\documentclass[a4paper]{article}
\usepackage{amsmath}
\usepackage{amsfonts}
\usepackage{amssymb}
\newtheorem{theorem}{Theorem}

\newtheorem{definition}[theorem]{Definition}

\newtheorem{proposition}[theorem]{Proposition}
\newenvironment{proof}[1][Proof]{\textbf{#1.} }{\ \rule{0.5em}{0.5em}}

\numberwithin{equation}{section}

\begin{document}

\title{\textbf{Generalized WDVV equations for }$F_{4}$\textbf{ pure N=2
Super-Yang-Mills theory}}
\author{ \\ \\ {\Large  \textbf{L.K. Hoevenaars, P.H.M. Kersten, R. Martini} } \\ \\ \\Department of Applied Mathematics, University of Twente\\P.O. Box 217, 7500 AE Enschede, The Netherlands \\ \\ }
\date{}
\maketitle

\begin{abstract}
An associative algebra of holomorphic differential forms is constructed
associated with pure N=2 Super-Yang-Mills theory for the Lie algebra $F_{4}$.
Existence and associativity of this algebra, combined with the general
arguments in the work of Marshakov, Mironov and Morozov, proves that the
prepotential of this theory satisfies the generalized WDVV system.
\end{abstract}

\section{Introduction}

In 1994, Seiberg and Witten \cite{SEIB-WITT1:1994}\ solved the low energy
behaviour of pure N=2 Super-Yang-Mills theory by giving the solution of the
prepotential $\mathcal{F}$. The essential ingredients in their construction
are a family of Riemann surfaces $\Sigma$, a meromorphic differential
$\lambda_{SW}$\ on it and the definition of the prepotential in terms of
period integrals of $\lambda_{SW}$%
\begin{equation}
a_{I}=\int_{A_{I}}\lambda_{SW}\text{ \ \ \ \ \ \ \ }\frac{\partial\mathcal{F}%
}{\partial a_{I}}=\int_{B_{I}}\lambda_{SW}%
\end{equation}
where $A_{I}$ and $B_{I}$ belong to a subset of the canonical cycles on the
surface $\Sigma$ and the $a_{I}$ are a subset of the moduli parameters of the
family of surfaces. These formulae define the prepotential $\mathcal{F}\left(
a_{1},...,a_{r}\right)  $ implicitly, where $r$ denotes the rank of the gauge
group under consideration.

A link between the prepotential and the Witten-Dijkgraaf-Verlinde-Verlinde
equations \cite{WITT:1991},\cite{DIJK-VERL-VERL:1991} was first suggested in
\cite{BONE-MATO:1996}. Since then an extensive literature on the subject was
formed. It was found that the perturbative piece of the prepotential
$\mathcal{F}(a_{1},...,a_{r})$ for pure N=2 SYM theory satisfies the
generalized WDVV equations \cite{MARS-MIRO-MORO:1996},\cite{MART-GRAG:1999}%
,\cite{VESE:1999}
\begin{equation}
\mathcal{F}_{I}\mathcal{F}_{K}^{-1}\mathcal{F}_{J}=\mathcal{F}_{J}%
\mathcal{F}_{K}^{-1}\mathcal{F}_{I}\text{ \ \ \ \ }\forall I,J,K=1,...,r
\end{equation}
where the $\mathcal{F}_{I}$ are matrices given by $\left(  \mathcal{F}%
_{I}\right)  _{JK}=\frac{\partial^{3}\mathcal{F}}{\partial a_{I}\partial
a_{J}\partial a_{K}}$.

Moreover it was shown that the full prepotential for classical Lie algebras
satisfies this generalized WDVV system \cite{MARS-MIRO-MORO:1996}%
,\cite{MARS-MIRO-MORO:2000},\cite{MARS-MIRO-MORO:1997}. The approach used by
these authors consists of constructing an associative algebra of holomorphic
differential forms, which together with a residue formula and existence of an
invertible metric proves that the prepotential satisfies the generalized WDVV
equations. For simply laced Lie algebras, an alternative proof was given in
\cite{ITO-YANG:1998}. Since these include the exceptional Lie algebras of type
$E_{6},E_{7},E_{8}$ this leaves the case of Lie algebra $F_{4}$
open\footnote{The algebra $G_{2}$ has rank two, which corresponds to two
variables $a_{1},a_{2}$. However, the generalized WDVV equations are trivial
for a function $\mathcal{F}$ of only two variables.}.

In this letter, we construct the algebra of differential forms for the Lie
algebra $F_{4}$ and we prove its associativity. Combined with the general
remarks of \cite{MARS-MIRO-MORO:2000},\cite{MARS-MIRO-MORO:1997} this proves
that the prepotential satisfies the WDVV equations.

\section{Associative algebra for $F_{4}$}

We start with the family of Riemann surfaces \cite{MART-WARN:1996}%
,\cite{ITO:1999} associated with pure $F_{4}$ Seiberg-Witten theory
\begin{equation}
z+\frac{\mu}{z}=W(x,u_{1},...,u_{4})={\frac{b_{1}(x)}{24}}-{\frac{1}{2}%
}\left\{  \left(  \frac{-q+\sqrt{q^{2}+4p^{3}}}{2}\right)  ^{1/3}+\left(
\frac{-q-\sqrt{q^{2}+4p^{3}}}{2}\right)  ^{1/3}\right\}  \label{definitie W}%
\end{equation}
where $p,q,b_{1}$ are polynomials in $x,u_{1},...,u_{4}$ which can be found in
Appendix \ref{appendix 1}. The Seiberg-Witten differential on this curve is
\begin{equation}
\lambda_{SW}=x\frac{dz}{z}=\frac{x\left(  \partial_{x}W\right)  dx}%
{\sqrt{W^{2}-4\mu}}%
\end{equation}
and its derivatives with respect to the moduli parameters $u_{i}$ are
holomorphic \cite{MARS-MIRO-MORO:1997}
\begin{equation}
\omega_{i}=\frac{\partial\lambda_{SW}}{\partial u_{i}}\cong-\frac{\partial
W}{\partial u_{i}}\frac{dx}{\sqrt{W^{2}-4\mu}}=\phi_{i}\frac{dx}{\sqrt
{W^{2}-4\mu}}%
\end{equation}
where $\cong$ denotes equality modulo exact forms and the last equality in
$\left(  \ref{phi}\right)  $ introduces the $\phi_{i}$. We want to make an
associative algebra out of a product structure for holomorphic differential forms
\begin{equation}
\omega_{i}\omega_{j}=\sum_{k=1}^{4}C_{ij}^{k}\omega_{k}G+H_{ij}\frac{dz}{z},
\label{algebra of forms}%
\end{equation}
where $G$ is a fixed holomorphic form and $H_{ij}$ are holomorphic forms.

There are mainly three circumstances that make the investigation different
from that of the classical algebras. For Lie algebras of type $A$, the study
of the algebra $\left(  \ref{algebra of forms}\right)  $ essentially comes
down to creating an algebra from a (commutative) ring of polynomials $\phi
_{i}\in\mathbb{C}\left[  x\right]  $ modulo the ideal generated by the fixed
polynomial $\partial_{x}W$. Such an algebra is automatically associative,
because any commutative ring modulo an ideal is again a commutative ring. For
the other classical Lie algebras, the $\phi_{i}$ and $\partial_{x}W$ need not
be polynomial, but a multiplication by $x^{\alpha}$ for some $\alpha$ makes
them polynomial and the same construction can be applied. In our case however
(as for all the exceptional groups), this strategy does not work due to the
cubic and square roots in $\left(  \ref{definitie W}\right)  $.

Furthermore, the Riemann surfaces do not have enough known involutions, which
facilitated the investigation for classical groups. These two problems will be
dealt with in the following sections: we will construct a polynomial algebra
in several variables and an involution of the Riemann surface is found.

Finally, the case of $F_{4}$ is more difficult from a purely computational
point of view, and calculations have to be done using a computer. We did
calculations in the symbolic languages REDUCE \cite{HEAR:1983} and MAPLE
\cite{MONA-GEDD-HEAL-LABA-VORK:1998}.

\subsection{The polynomial ring and ideals}

Due to the cubic and square roots in (\ref{definitie W}), the $\phi_{i}%
=-\frac{\partial W}{\partial u_{i}}$ have terms containing $\left(
-q+\sqrt{q^{2}+4p^{3}}\right)  ^{-\frac{2}{3}}$ which are certainly not
polynomial in $x$. For classical gauge groups this problem does not occur and
the $\phi_{i}$ are basically in a polynomial ring. It is desirable to work
with a polynomial ring because it will lead to associativity of the algebra
structure (\ref{algebra of forms}). For this purpose we set
\begin{align}
c  &  =\sqrt{q^{2}+4p^{3}}\nonumber\\
a  &  =p\left(  \frac{-q+c}{2}\right)  ^{1/3}\\
b  &  =p\left(  \frac{-q-c}{2}\right)  ^{1/3}\nonumber
\end{align}
and $\widetilde{\phi}_{i}:=abc\phi_{i}$. With these definitions the Riemann
surface reads
\begin{equation}
z+\frac{\mu}{z}={\frac{b_{1}(x)}{24}}-{\frac{1}{2p}}\left(  a+b\right)
\end{equation}
The $\widetilde{\phi}_{i}$ are polynomial not in one variable $x$, but in four
variables $x,a,b,c$:

\begin{proposition}
\label{proof in appendix}The $\widetilde{\phi}_{i}$ are elements of the
polynomial ring $\mathbb{C}\left[  x,a,b,c\right]  $.
\end{proposition}

\begin{proof}
See appendix \ref{appendix proof}.
\end{proof}

Due to the definitions of $a,b,c$ there are certain relations among them. When
multiplying the $\widetilde{\phi}_{i}$ to obtain an algebra, we have to take
into account that
\begin{align}
c^{2}-q^{2}-4p^{3}  &  =0\tag{I.1}\\
ab+p^{3}  &  =0\tag{I.2}\\
a^{2}-\frac{1}{2}\left(  q-c\right)  b  &  =0\tag{I.3}\\
b^{2}-\frac{1}{2}\left(  q+c\right)  a  &  =0 \tag{I.4}%
\end{align}
These equations generate also other polynomial relations between $a,b,c$. For
example, from the definition of $a$ it is clear that $a^{3}=\frac{1}{2}\left(
-q+c\right)  p^{3}$. This relation can also be deduced from $(I.3)$ and
$(I.2):$%
\begin{equation}
a^{3}=a\cdot a^{2}=ab\frac{1}{2}\left(  q-c\right)  =\frac{1}{2}\left(
-q+c\right)  p^{3}%
\end{equation}
We will make practical use of these relations via the following

\begin{definition}
The equations (I.1),(I.2),(I.3),(I.4) generate\footnote{Note that defining
$\widetilde{a}=\left(  \frac{-q+c}{2}\right)  ^{1/3}$ would yield
$\widetilde{a}^{2}-\frac{1}{2p}\left(  qb-bc\right)  =0$ which is
\underline{not} polynomial. This is the reason why an extra factor $p$ is
added in the definition of $a$ and $b$.} an ideal $I$ in $\mathbb{C}\left[
x,a,b,c\right]  $.
\end{definition}

So in fact the $\widetilde{\phi}_{i}$ are in $\mathbb{C}\left[
x,a,b,c\right]  /I$ and for any equivalence class $p(x,a,b,c)+I$ of this space
we can take a representative of the following form:
\begin{equation}
p(x,a,b,c)+I=p_{1}(x)+ap_{2}(x)+bp_{3}(x)+cp_{4}(x)+acp_{5}(x)+bcp_{6}(x)+I
\end{equation}
because any higher powers of $a,b,c$ can be rewritten using the ideal.

Since we expect the $H_{ij}$ to be holomorphic differentials, we have taken
the Ansatz that they are of the same form as the $\omega_{i}$. Since
\begin{equation}
\omega_{i}=\phi_{i}\frac{dx}{\sqrt{W^{2}-4\mu}}=\frac{1}{abc}\widetilde{\phi
}_{i}\frac{dx}{\sqrt{W^{2}-4\mu}}%
\end{equation}
with $\phi_{i}$ as in $(\ref{vorm phi})$, we take
\begin{align}
H_{ij}  &  =Q_{ij}\frac{dx}{\sqrt{W^{2}-4\mu}}=\frac{1}{abc}\widetilde{Q}%
_{ij}\frac{dx}{\sqrt{W^{2}-4\mu}}\nonumber\label{maximal degree}\\
&  =\frac{1}{abc}\left(  abp_{ij1}(x)+ap_{ij2}(x)+bp_{ij3}(x)+abcp_{ij4}%
(x)+acp_{ij5}(x)+bcp_{ij6}(x)\right)  \frac{dx}{\sqrt{W^{2}-4\mu}}%
\end{align}
With this choice, the classes represented by $\widetilde{Q}_{ij}$ are elements
of $\mathbb{C}\left[  x,a,b,c\right]  /I$. Final part of the Ansatz is that
the polynomials $p_{ijk}(x)$ are graded in the variables $(x,u_{i})$ with a
certain degree which will be determined in section \ref{section grading}.

\bigskip

Furthermore we have to make a choice for the holomorphic differential $G$. In
the $ADE$ cases, there are reasons \cite{ITO-YANG:1998} to take $G=\omega
_{r}=\frac{dx}{\sqrt{W^{2}-4\mu}}$ where $r$ is the rank of the group. By
analogy we take $G=\omega_{4}$.

Instead of the multiplication structure $\left(  \ref{algebra of
forms}\right)  $ we now look at the equivalent structure in local coordinates
\begin{equation}
\phi_{i}\phi_{j}=\sum_{k=1}^{4}C_{ij}^{k}\phi_{k}\phi_{4}+Q_{ij}\partial_{x}W
\label{algebra of phi's}%
\end{equation}
In terms of the $\widetilde{\phi}_{i}$ we get
\begin{equation}
\left(  \frac{1}{abc}\right)  ^{2}\widetilde{\phi}_{i}\widetilde{\phi}%
_{j}=\left(  \frac{1}{abc}\right)  ^{2}\left(  \sum_{k=1}^{4}C_{ij}%
^{k}\widetilde{\phi}_{k}\widetilde{\phi}_{4}+\widetilde{Q}_{ij}\widetilde
{\partial_{x}W}\right)  \label{multiplication with abc}%
\end{equation}
where $\widetilde{\partial_{x}W}=abc\partial_{x}W$. Equation
$(\ref{multiplication with abc})$ is equivalent to
\begin{equation}
\widetilde{\phi}_{i}\widetilde{\phi}_{j}=\sum_{k=1}^{4}C_{ij}^{k}%
\widetilde{\phi}_{k}\widetilde{\phi}_{4}+\widetilde{Q}_{ij}\widetilde
{\partial_{x}W} \label{structure}%
\end{equation}
in $\mathbb{C}\left[  x,a,b,c\right]  /I$. From this multiplication structure,
the existence of which will be discussed in section \ref{existence}, we can
construct an algebra

\begin{definition}
We define the algebra $A$ by defining the multiplication $\ast:\mathbb{C}%
\left[  x,a,b,c\right]  /I\times\mathbb{C}\left[  x,a,b,c\right]
/I\rightarrow\mathbb{C}\left[  x,a,b,c\right]  /I$ by
\begin{equation}
\widetilde{\phi}_{i}\ast\widetilde{\phi}_{j}=\sum_{k=1}^{4}C_{ij}%
^{k}\widetilde{\phi}_{k}%
\end{equation}
where the structure constants are taken from (\ref{structure}). $\widetilde
{\phi}_{4}$ is the unity for this multiplication.
\end{definition}

The ideal $I$ and the ideal generated by $abc\partial_{x}W$ together give a
new ideal $J$\ in $\mathbb{C}\left[  x,a,b,c\right]  $. It can be shown that
$\mathbb{C}\left[  x,a,b,c\right]  /J$ is finite dimensional. The algebra $A$
is obtained from polynomial multiplication modulo this ideal and it yields a
4-dimensional subalgebra of $\mathbb{C}\left[  x,a,b,c\right]  /J$. By our
construction we have proven associativity:

\begin{theorem}
\label{the theorem}The algebra $A$ is associative.
\end{theorem}

With this result, the problem of finding an appropriate polynomial ring has
been overcome. The following section deals with a symmetry and grading of the problem.

\subsection{Symmetries and grading\label{section grading}}

A very important tool in calculations is the grading which is present in the
problem (see for example \cite{MART-WARN:1996}). The origin of this grading
lies in a grading of the underlying Lie algebra. We will list the degrees:
\[%
\begin{tabular}
[c]{lllll}%
$\left[  x\right]  =1$ \ \ , \ \  & $\left[  u_{1}\right]  =2$ \ \ , \ \  &
$\left[  u_{2}\right]  =6$ \ \ , \ \  & $\left[  u_{3}\right]  =8$ \ \ ,
\ \  & $\left[  u_{4}\right]  =12$%
\end{tabular}
\]
and from this grading we can deduce the degrees of all other objects (see
appendix \ref{appendix 1}). For example, $\left[  \phi_{1}\right]  =7$ so
$\left[  Q_{11}\right]  =\left[  \phi_{1}\phi_{1}\right]  -\left[
\partial_{x}W\right]  =14-8=6$ and from this we can deduce the degrees of the
$p_{11k}(x)$ of equation $\left(  \ref{maximal degree}\right)  $ as promised.
For $H_{11}$ we get
\[%
\begin{tabular}
[c]{llllll}%
$\left[  p_{111}\right]  =33$ \ \ , \ \  & $\left[  p_{112}\right]  =60$ \ \ ,
\ \  & $\left[  p_{113}\right]  =60$ \ \ , \ \  & $\left[  p_{114}\right]  =6$
\ \ , \ \  & $\left[  p_{115}\right]  =33$ \ \ , \ \  & $\left[
p_{116}\right]  =33$%
\end{tabular}
\]

Apart from the involution $z\rightarrow\frac{\mu}{z}$ which is known to exist
for all Riemann surfaces associated with pure Seiberg-Witten
theory\cite{MART-WARN:1996}, there is at least one other involution present
for $F_{4}$. As a result of the grading, we have the following involution of
the Riemann surface: $z\rightarrow-z$ , \ $x\rightarrow-x$. This involution is
also exhibited by the surfaces of $B_{r},C_{r}$ Seiberg-Witten theory, which
are constructed from the same type of procedure \cite{MART-WARN:1996}.

Under a rescaling of $x$ by a factor $\alpha$, the other objects must
transform according to their degree%
\[%
\begin{tabular}
[c]{llllll}%
$x\rightarrow\alpha x$ \ \ , \ \  & $z\rightarrow\alpha^{9}z$ \ \ , \ \  &
$u_{1}\rightarrow\alpha^{2}u_{1}$ \ \ , \ \  & $u_{2}\rightarrow\alpha
^{6}u_{2}$ \ \ , \ \  & $u_{3}\rightarrow\alpha^{8}u_{3}$ \ \ , \ \  &
$u_{4}\rightarrow\alpha^{12}u_{4}$%
\end{tabular}
\]
and substituting $\alpha=-1$ gives the involution. The reason for this
symmetry is therefore that $x$ and $z$ have odd degrees, whereas the Casimirs
$u_{i}$ all have even degrees. Under this symmetry, the objects transform as
follows%
\[%
\begin{tabular}
[c]{lll}%
$W\mapsto-W$ & \ \ \ \  & $a\mapsto-b$\\
$\phi_{i}\mapsto-\phi_{i}$ & \ \ \ \  & $b\mapsto-a$\\
$\partial_{x}W\mapsto\partial_{x}W$ & \ \ \ \  & $c\mapsto c$%
\end{tabular}
\]
Using this symmetry in the multiplication structure $\left(  \ref{algebra of
phi's}\right)  $, we find that $Q_{ij}\mapsto Q_{ij}$. This facilitates the
computations by narrowing down the possible forms $H_{ij}$.

\subsection{Construction of the algebra\label{existence}\label{structure
constants}}

The procedure we have used to calculate the multiplication structure $\left(
\ref{structure}\right)  $ is to set each coefficient of the multivariate
polynomial in $x,a,b,c,u_{1},...,u_{4}$ of the left hand side equal to that of
the right hand side. This yields an overdetermined system of linear equations
which can be solved uniquely to give the following structure
constants\footnote{To get a better lay-out, we give the transpose matrices
$\left(  C_{i}^{T}\right)  _{j}^{k}=\left(  C_{i}\right)  _{k}^{j}$.}:

\bigskip

$\left(  C_{1}^{T}\right)  _{j}^{k}=\left(
\begin{array}
[c]{cccc}%
u_{1}\left(  \frac{250}{243}u_{1}^{4}-\frac{10}{9}u_{1}u_{2}-\frac{7}{3}%
u_{3}\right)  & -\frac{25}{54}u_{1}^{3}+\frac{1}{4}u_{2} & -\frac{5}{3}%
u_{1}^{2} & 1\\
&  &  & \\
\frac{100}{81}u_{1}^{4}u_{2}+\frac{140}{27}u_{1}^{3}u_{3}-\frac{2}{3}%
u_{1}u_{2}^{2}-\frac{4}{3}u_{1}u_{4}-2u_{2}u_{3} & u_{1}\left(  -\frac{5}%
{9}u_{1}u_{2}-\frac{7}{3}u_{3}\right)  & -6u_{3}-2u_{1}u_{2} & 0\\
&  &  & \\
-\frac{2}{9}u_{1}u_{2}u_{3}-\frac{2}{3}u_{3}^{2}+\frac{100}{243}u_{1}^{4}%
u_{3}-\frac{10}{27}u_{1}^{2}u_{4} & \frac{1}{6}u_{4}-\frac{5}{27}u_{1}%
^{2}u_{3} & -\frac{2}{3}u_{1}u_{3} & 0\\
&  &  & \\
\frac{10}{9}u_{1}^{2}u_{3}^{2}-\frac{1}{3}u_{1}u_{2}u_{4}-u_{3}u_{4}+\frac
{50}{81}u_{1}^{4}u_{4} & -\frac{1}{2}u_{3}^{2}-\frac{5}{18}u_{1}^{2}u_{4} &
-u_{1}u_{4} & 0
\end{array}
\right)  $

{\tiny \bigskip}

$\left(  C_{2}^{T}\right)  _{j}^{k}=\left(
\begin{array}
[c]{cccc}%
-\frac{25}{54}u_{1}^{3}+\frac{1}{4}u_{2} & \frac{5}{24}u_{1} & \frac{3}{4} &
0\\
&  &  & \\
u_{1}\left(  -\frac{5}{9}u_{1}u_{2}-\frac{7}{3}u_{3}\right)  & \frac{1}%
{4}u_{2} & 0 & 1\\
&  &  & \\
\frac{1}{6}u_{4}-\frac{5}{27}u_{1}^{2}u_{3} & \frac{1}{12}u_{3} & 0 & 0\\
&  &  & \\
-\frac{1}{2}u_{3}^{2}-\frac{5}{18}u_{1}^{2}u_{4} & \frac{1}{8}u_{4} & 0 & 0
\end{array}
\right)  $

{\tiny \bigskip}

$\left(  C_{3}^{T}\right)  _{j}^{k}=\left(
\begin{array}
[c]{cccc}%
-\frac{5}{3}u_{1}^{2} & \frac{3}{4} & 0 & 0\\
&  &  & \\
-6u_{3}-2u_{1}u_{2} & 0 & -6u_{1} & 0\\
&  &  & \\
-\frac{2}{3}u_{1}u_{3} & 0 & 0 & 1\\
&  &  & \\
-u_{1}u_{4} & 0 & -\frac{9}{2}u_{3} & 0
\end{array}
\right)  $

{\tiny \bigskip}

$\left(  C_{4}^{T}\right)  _{j}^{k}=\left(
\begin{array}
[c]{cccc}%
1 & 0 & 0 & 0\\
0 & 1 & 0 & 0\\
0 & 0 & 1 & 0\\
0 & 0 & 0 & 1
\end{array}
\right)  $

\bigskip

Although it was proven abstractly in theorem \ref{the theorem}\ that these are
structure constants of an associative algebra, this was also checked
explicitly from the expressions above.

\section{Conclusions and Outlook}

In this letter, we constructed the algebra of holomorphic differential forms
for Lie algebra $F_{4}$ and we proved its associativity. Together with the
theory of \cite{MARS-MIRO-MORO:2000} this proves that the
prepotential of pure $F_{4}$ Seiberg-Witten theory satisfies the generalized
WDVV equations. Apart from the link of Seiberg-Witten theory with integrable
systems \cite{GORS-KRIC-MARS-MIRO-MORO:1995},\cite{MART-WARN:1996} there is no
explanation why the generalized WDVV equations should hold for this theory.
One possible indication is that its origin lies in the 2D Landau-Ginzburg
systems (for which the WDVV equations themselves hold), as explained for
simply laced groups in \cite{ITO-YANG:1998}. In that article it was also
conjectured that for $B,C$ type Lie algebras, the generalized WDVV equations
can be shown to hold by using the Landau-Ginzburg theory of $BC$ type. This
was subsequently proven in \cite{futurework}. It would be interesting to use
the algebra constructed explicitly in the present paper for $F_{4}$ to find
out if an interpretation of the generalized WDVV equations in terms of the
$F_{4}$ Landau-Ginzburg model \cite{ZUBE:1994}\ can be given.

\bigskip

\textbf{Acknowledgements:} It is a pleasure to thank G. Post and G. Helminck
for stimulating discussions.

\bigskip

\bigskip\appendix

\section{\label{appendix 1}The $F_{4}$ spectral curve}

The $F_{4}$ spectral curve\footnote{Note that we have corrected a misprint in
\cite{ITO:1999} by adding a factor $\frac{1}{2}$ in the cube roots.} is given
by (\cite{MART-WARN:1996},\cite{ITO:1999})
\begin{equation}
z+\frac{\mu}{z}=W(x,u_{1},...,u_{4})={\frac{b_{1}(x)}{24}}-{\frac{1}{2}%
}\left\{  \left(  \frac{-q+\sqrt{q^{2}+4p^{3}}}{2}\right)  ^{1/3}+\left(
\frac{-q-\sqrt{q^{2}+4p^{3}}}{2}\right)  ^{1/3}\right\}
\end{equation}
where
\begin{align}
p(x)  &  =-{\frac{b_{2}}{6}}-{\frac{b_{1}^{2}}{144}},\nonumber\\
q(x)  &  ={\frac{1}{27}}\left(  {\frac{b_{1}^{3}}{32}}+{\frac{9}{8}}b_{1}%
b_{2}+27b_{3}\right)
\end{align}
and
\begin{align}
b_{1}(x)  &  =-636x^{9}-300{u_{1}}x^{7}-48{u_{1}^{2}}x^{5}-5{u_{2}}%
x^{3}+2{u_{3}}x,\nonumber\\
b_{2}(x)  &  =-168x^{18}-348{u_{1}}x^{16}-276u_{1}^{2}x^{14}+(-116u_{1}%
^{3}+14{u_{2}})x^{12}\nonumber\\
&  +(-92{u_{3}}-20u_{1}^{4}-8{u_{1}u_{2}})x^{10}+(-42{u_{1}u_{3}}-6u_{1}%
^{2}{u_{2}})x^{8}\nonumber\\
&  +(-4{u_{4}}-{\frac{10}{3}}u_{1}^{2}{u_{3}}-{\frac{2}{3}}u_{2}^{2}%
)x^{6}+({\frac{1}{3}u_{2}u_{3}}-{\frac{2}{3}u_{4}u_{1}})x^{4},\nonumber\\
b_{3}(x)  &  =x^{27}+6{u_{1}}x^{25}+15u_{1}^{2}x^{23}+(20u_{1}^{3}+{u_{2}%
})x^{21}+(5{u_{3}}+4{u_{1}u_{2}}+15u_{1}^{4})x^{19}\nonumber\\
&  +(6u_{1}^{2}{u_{2}}+12{u_{1}u_{3}}+6{u_{1}^{5}})x^{17}+({\frac{1}{3}}%
u_{2}^{2}+5{u_{4}}+4u_{1}^{3}{u_{2}}+{\frac{26}{3}}u_{1}^{2}{u_{3}}+{u_{1}%
^{6}})x^{15}\nonumber\\
&  +({\frac{4}{3}}u_{1}^{3}{u_{3}}+{\frac{19}{3}u_{4}u_{1}}+u_{1}^{4}{u_{2}%
}+{\frac{4}{3}u_{2}u_{3}}+{\frac{2}{3}}u_{2}^{2}{u_{1}})x^{13}\nonumber\\
&  +({\frac{1}{3}}u_{1}^{2}u_{2}^{2}-{\frac{1}{3}}u_{1}^{4}{u_{3}}-{\frac
{15}{4}}u_{3}^{2}+3{u_{4}}u_{1}^{2})x^{11}\nonumber\\
&  +({\frac{1}{3}u_{4}u_{2}}-{\frac{4}{9}}u_{1}^{2}{u_{2}u_{3}}+{\frac{1}{27}%
}u_{2}^{3}-{\frac{13}{6}}u_{3}^{2}{u_{1}}+{\frac{13}{27}u_{4}}u_{1}^{3}%
)x^{9}\nonumber\\
&  +(-{\frac{1}{9}}u_{2}^{2}{u_{3}}-{\frac{1}{2}u_{4}u_{3}}+{\frac{1}{9}%
u_{4}u_{1}u_{2}}-{\frac{7}{36}}u_{1}^{2}u_{3}^{2})x^{7}+({\frac{1}{12}}%
u_{3}^{2}{u_{2}}-{\frac{1}{6}u_{4}u_{1}u_{3}})x^{5}\nonumber\\
&  +(-{\frac{1}{54}}u_{3}^{3}-{\frac{1}{108}u_{4}^{2}})x^{3}.
\end{align}

\bigskip

The degrees of several objects, induced by the grading in section \ref{section
grading}, are given in the following table:
\[%
\begin{tabular}
[c]{|c|c|c|c|c|c|c|c|c|c|c|}\hline
$\left[  p\right]  $ & $\left[  q\right]  $ & $\left[  a\right]  $ & $\left[
b\right]  $ & $\left[  c\right]  $ & $\left[  W\right]  $ & $\left[
\partial_{x}W\right]  $ & $\left[  \phi_{1}\right]  $ & $\left[  \phi
_{2}\right]  $ & $\left[  \phi_{3}\right]  $ & $\left[  \phi_{4}\right]
$\\\hline
$18$ & $27$ & $27$ & $27$ & $27$ & $9$ & $8$ & $7$ & $3$ & $1$ & $-3$\\\hline
\end{tabular}
\]

\section{\label{appendix proof}Proof of proposition \ref{proof in appendix}}

The $\phi_{i}$ are of the form
\begin{align}
\phi_{i}  &  =-\frac{\partial W}{\partial u_{i}}=-\frac{\partial}{\partial
u_{i}}\left(  \frac{b_{1}}{24}-\frac{1}{2p}\left(  a+b\right)  \right)
=-\frac{1}{24}\frac{\partial b_{1}}{\partial u_{i}}-\frac{1}{2p^{2}}\left(
a+b\right)  +\frac{1}{2p}\frac{\partial}{\partial u_{i}}\left(  a+b\right)
\nonumber\\
&  =-\frac{1}{24}\frac{\partial b_{1}}{\partial u_{i}}-\frac{1}{2p^{2}}\left(
a+b\right)  +\frac{1}{2p^{2}}\left(  a+b\right)  +\nonumber\\
&  \text{ \ \ \ \ \ \ \ \ \ \ \ \ \ }\frac{p^{2}}{6a^{2}}\left(  -\frac{1}%
{2}\frac{\partial q}{\partial u_{i}}+\frac{1}{4c}\frac{\partial}{\partial
u_{i}}\left(  q^{2}+4p^{3}\right)  \right)  +\frac{p^{2}}{6b^{2}}\left(
-\frac{1}{2}\frac{\partial q}{\partial u_{i}}-\frac{1}{4c}\frac{\partial
}{\partial u_{i}}\left(  q^{2}+4p^{3}\right)  \right) \nonumber\\
&  =-\frac{1}{24}\frac{\partial b_{1}}{\partial u_{i}}+\frac{p^{3}}{6a^{2}%
}\left(  -\frac{1}{2}\frac{\partial q}{\partial u_{i}}+\frac{q}{2c}%
\frac{\partial q}{\partial u_{i}}+\frac{3p^{2}}{c}\frac{\partial p}{\partial
u_{i}}\right)  +\frac{p^{3}}{6b^{2}}\left(  -\frac{1}{2}\frac{\partial
q}{\partial u_{i}}-\frac{q}{2c}\frac{\partial q}{\partial u_{i}}-\frac{3p^{2}%
}{c}\frac{\partial p}{\partial u_{i}}\right)  \label{phi}%
\end{align}
Now we use the relations $\left(  I.1\right)  -\left(  I.4\right)  $ to see
that
\begin{align}
\frac{1}{a^{2}}  &  =\frac{2}{b\left(  q-c\right)  }=\frac{2\left(
q+c\right)  }{b\left(  q-c\right)  \left(  q+c\right)  }=\frac{-\left(
q+c\right)  }{2bp^{3}}=-\frac{q}{2p^{3}b}-\frac{q^{2}+4p^{3}}{2p^{3}%
bc}\nonumber\\
\frac{1}{a^{2}c}  &  =\frac{1}{c}\left(  -\frac{q}{2p^{3}b}-\frac{q^{2}%
+4p^{3}}{2p^{3}bc}\right)  =-\frac{q}{2p^{3}bc}-\frac{1}{2p^{3}b}%
\end{align}
In a similar way one derives
\begin{align}
\frac{1}{b^{2}}  &  =\frac{-q}{2p^{3}a}+\frac{q^{2}+4p^{3}}{2p^{3}%
ac}\nonumber\\
\frac{1}{b^{2}c}  &  =-\frac{q}{2p^{3}ac}+\frac{1}{2p^{3}a}%
\end{align}
Using these equations in $\left(  \ref{phi}\right)  $\ yields
\begin{gather}
\phi_{i}=-\frac{1}{24}\frac{\partial b_{1}}{\partial u_{i}}+\frac{1}{6}\left(
-\frac{q}{2b}-\frac{q^{2}+4p^{3}}{2bc}\right)  \left(  -\frac{1}{2}%
\frac{\partial q}{\partial u_{i}}\right)  +\frac{1}{6}\left(  -\frac{q}%
{2bc}-\frac{1}{2b}\right)  \left(  \frac{q}{2}\frac{\partial q}{\partial
u_{i}}+3p^{2}\frac{\partial p}{\partial u_{i}}\right) \nonumber\\
+\frac{1}{6}\left(  -\frac{q}{2a}+\frac{q^{2}+4p^{3}}{2ac}\right)  \left(
-\frac{1}{2}\frac{\partial q}{\partial u_{i}}\right)  +\frac{1}{6}\left(
-\frac{q}{2ac}+\frac{1}{2a}\right)  \left(  -\frac{q}{2}\frac{\partial
q}{\partial u_{i}}-3p^{2}\frac{\partial p}{\partial u_{i}}\right)
\label{vorm phi}%
\end{gather}
and therefore the $\widetilde{\phi}_{i}=abc\phi_{i}$ are polynomials in
$x,a,b,c$.

\bibliographystyle{h-physrev}
\bibliography{biblio}

\begin{thebibliography}{10}

\bibitem{SEIB-WITT1:1994}
N.~Seiberg and E.~Witten,
\newblock Nucl. Phys. {\bf B426}, 19 (1994), hep-th/9407087.

\bibitem{WITT:1991}
E.~Witten,
\newblock Two-dimensional gravity and intersection theory on moduli space,
\newblock in {\em Surveys in differential geometry (Cambridge, MA, 1990)}, pp.
  243--310, Lehigh Univ., Bethlehem, PA, 1991.

\bibitem{DIJK-VERL-VERL:1991}
R.~Dijkgraaf, H.~Verlinde, and E.~Verlinde,
\newblock Nucl. Phys. {\bf B352}, 59 (1991).

\bibitem{BONE-MATO:1996}
G.~Bonelli and M.~Matone,
\newblock Phys. Rev. Lett. {\bf 77}, 4712 (1996), hep-th/9605090.

\bibitem{MARS-MIRO-MORO:1996}
A.~Marshakov, A.~Mironov, and A.~Morozov,
\newblock Phys. Lett. {\bf B389}, 43 (1996), hep-th/9607109.

\bibitem{MART-GRAG:1999}
R.~Martini and P.~K.~H. Gragert,
\newblock J. Nonlinear Math. Phys. {\bf 6}, 1 (1999).

\bibitem{VESE:1999}
A.~P. Veselov,
\newblock Phys. Lett. {\bf A261}, 297 (1999), hep-th/9902142.

\bibitem{MARS-MIRO-MORO:2000}
A.~Marshakov, A.~Mironov, and A.~Morozov,
\newblock Int. J. Mod. Phys. {\bf A15}, 1157 (2000), hep-th/9701123.

\bibitem{MARS-MIRO-MORO:1997}
A.~Marshakov, A.~Mironov, and A.~Morozov,
\newblock Mod. Phys. Lett. {\bf A12}, 773 (1997), hep-th/9701014.

\bibitem{ITO-YANG:1998}
K.~Ito and S.-K. Yang,
\newblock Phys. Lett. {\bf B433}, 56 (1998), hep-th/9803126.

\bibitem{MART-WARN:1996}
E.~Martinec and N.~Warner,
\newblock Nucl. Phys. {\bf B459}, 97 (1996), hep-th/9509161.

\bibitem{ITO:1999}
K.~Ito,
\newblock Prog. Theor. Phys. Suppl. {\bf 135}, 94 (1999), hep-th/9906023.

\bibitem{HEAR:1983}
A.~Hearn,
\newblock {\em {R}{E}{D}{U}{C}{E} user's manual (version 3.6)} (The Rand
  corporation, Santa Monica, 1983).

\bibitem{MONA-GEDD-HEAL-LABA-VORK:1998}
M.~Monagan, K.~Geddes, K.~Heal, G.~Labahn, and S.~Vorkoetter,
\newblock {\em Maple {V} programming guide for release 5} (Springer Verlag, New
  York, 1998).

\bibitem{GORS-KRIC-MARS-MIRO-MORO:1995}
A.~Gorsky, I.~Krichever, A.~Marshakov, A.~Mironov, and A.~Morozov,
\newblock Phys. Lett. {\bf B355}, 466 (1995), hep-th/9505035.

\bibitem{futurework}
L.~Hoevenaars and R.~Martini,
\newblock In preparation .

\bibitem{ZUBE:1994}
J.~B. Zuber,
\newblock Mod. Phys. Lett. {\bf A9}, 749 (1994), hep-th/9312209.

\end{thebibliography}
\end{document}